\documentclass[prl,aps,twocolumn,groupedaddress]{revtex4}
\usepackage{epsfig,latexsym,amssymb,amsmath,amsbsy,graphics,graphicx}

\def \beq{\begin{equation}}
\def \eeq{\end{equation}}
\def \beqarr{\begin{eqnarray}}
\def \eeqarr{\end{eqnarray}}

\begin{document}

\title{Phase Diagrams of Spinor Bose Gases
}

\author{Kun Yang}
\affiliation{NHMFL and Department of Physics, Florida State
University, Tallahassee, Florida 32306, USA}

\date{\today}

\begin{abstract}

Using effective field theories dictated by the symmetry of the system, as well as microscopic considerations, we map out the magnetic coupling-temperature phase diagrams of spin-1 Bose gases in both two- and three-dimensions. We also determine the nature of all phase boundaries, and critical properties in the case of 2nd order phase transitions at both zero and finite temperatures.

\end{abstract}

\maketitle

Spinor Bose condensates have been of strong interest for over a decade\cite{experiment,ho,ohmi,law}. Due to the presence of hyperfine degrees of freedom, such condensates have very rich internal structures and support a variety of novel topological defects\cite{ho,zhou,zhoureview}. In particular, depending on the values of two-particle s-wave scattering lengths in the singlet ($S=0$) and quintuplet ($S=2$) channels, $a_0$ and $a_2$, the system can form either nematic (polar) or ferromagnetic spinor condensates\cite{ho}. In the meantime effects of thermal fluctuations are also much stronger in these systems, and it was shown recently\cite{moore} that the nematic (or polar) spinor condensate of spin-1 atoms is unstable at any finite temperature in two-dimension (2D), yielding to a pair condensate with power-law superfluid order until a Kosterlitz-Thouless (KT) transition temperature is reached. ``Pairing" here is due to the thermal disordering of the nematic order parameter, instead of formation of bound states between atoms, and the resultant state is found to have topological spin order\cite{song}. Above the KT temperature the system loses superfluidity altogether. Related work\cite{podolsky} has focused on the case with easy-plane anisotropy, also in 2D, and found a multitude of KT transitions. Comparatively speaking, there has been much less discussion on the fate of ferromagnetic condensates, as well as what happens in three-dimension (3D).

In the present work we clarify the phase diagrams of spin-1 Bose gases with spin-rotation symmetry, in both 2D and 3D, and treat the nematic and ferromagnetic condensates on equal footing. Our results are summarized in two phase diagrams, Fig. 1 and Fig. 2, for the 2D and 3D cases respectively. Of crucial importance in our consideration are the existence of a high-symmetry boundary that separates the nematic and ferromagnetic phases, and a quantum critical point separating the nematic condensate state from a pair condensate state. The rest of the paper will be devoted to elaborating on the physics associated with the various phases and phase transitions in these two phase diagrams. We will also briefly comment on what happens in one-dimension (1D) toward the end.

The microscopic Hamiltonian describing spin-1 atoms with short-range interaction takes the form\cite{ho}:
\beqarr
H&=&\int{d{\bf r}}\left[{\hbar^2\over 2M}\nabla\psi_a^\dagger\cdot\nabla\psi_a+{c_0\over 2}\psi_a^\dagger \psi_b^\dagger \psi_b\psi_a \right. \nonumber\\
&&\left.+{c_2\over 2}\psi_a^\dagger \psi_{a'}^\dagger{\bf F}_{ab}\cdot {\bf F}_{a'b'} \psi_{b'}\psi_{b}\right],
\label{hamiltonian}
\eeqarr
where $\psi$ is the boson field, $M$ is atom mass, $a,a',b,b'$ are spin indices running from $-1$ to $+1$, the vector ${\bf F}$ has three components which are the 3 spin-1 matrices, and repeated indices are summed over. $c_0$ and $c_2$ are two-atom s-wave interaction constants; specifically, $c_0-2c_2$ and $c_0+c_2$ correspond to the interaction strength in the total spin $0$ and $2$ channels respectively\cite{g}. Stability of the system requires $c_0 > 0$, and $c_0 > |c_2|$ when $c_2 < 0$, although the actual value of $c_0$ is not very important in determining which phase the system is in; we thus simply assume it takes some sufficiently large value such that the two-body interaction is always repulsive in the quintuplet channel (thus $a_2 > 0$) and do not discuss it further. We have also neglected the trap potential for simplicity, and assume instead the system is translationally invariant and in the thermodynamic limit. The Hamiltonian (\ref{hamiltonian}) possesses an U(1)$\times$SU(2) symmetry, where U(1) corresponds to charge conservation, and SU(2) corresponds to spin-rotation symmetry.

The case with $c_2=0$ is very special. In this case the interaction, and therefore the entire Hamiltonian (\ref{hamiltonian}) is {\em spin-independent}. As a result the Hamiltonian (\ref{hamiltonian}) has a much higher U(1)$\times$SU(3) symmetry\cite{lieb}, which one can understand by treating the 3 internal spin states as the fundamental representation of the SU(3) group. This case is of special importance in our discussion below. Experimentally, this corresponds to the case that the scattering lengthes in the singlet and quintuplet channels are the same: $a_0=a_2>0$.

The spin structure of the spinor condensate is determined by $c_2$\cite{ho}. For $c_2 < 0$ it is energetically favorable to have {\em all} atom spins aligned ferromagnetically; the ground state is a fully magnetized ferromagnet. For $c_2 > 0$, spin polarization is suppressed and nematic spin states become energetically favorable. In particular,
as we show below,
when $c_2$ becomes big enough, singlet bound states or molecules may form, and spin order disappears.

\begin{centering}
\begin{figure}
\epsfig {file=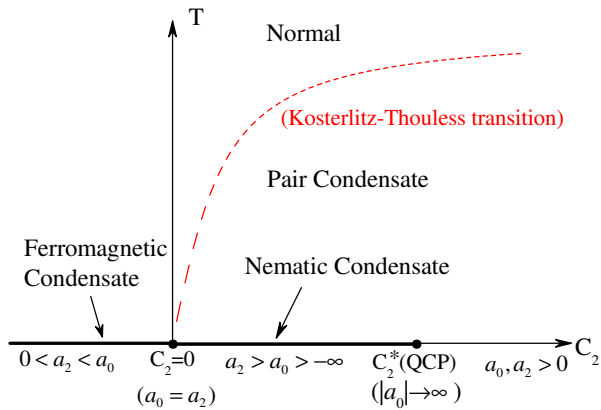,width=80mm}
\caption{(Color online)
Phase diagram of a spin-1 Bose gas in 2D. At $T=0$ there are three phases: Ferromagnetic Condensate, Nematic Condensate and Pair Condensate, separated by quantum phase transitions at $c_2=0$ and $c_2=c_2^*$ respectively. At finite $T$ only the Pair Condensate phase survives, up to a Kosterlitz-Thouless transition, above which the system is in a normal phase with no broken symmetry.
}
\label{fig1}
\end{figure}
\end{centering}

{\em Phases and Quantum Phase Transitions at Zero Temperature} -- We start by considering the phases and phase transitions at $T=0$, where our analysis applies to both 3D and 2D. We expect 3 different phases.

(i) Ferromagnetic condensate (FMC) phase with $c_2 < 0$, where $0 < a_2 < a_0$. In this phase atoms condense into a spin-polarized state:
$\zeta_{FMC}=e^{i\theta}(1, 0, 0)^T$,
up to an arbitrary global rotation and gauge transformation. The order parameter is parameterized by $({\bf n}, e^{i\theta})$, where $\theta$ is the superfluid phase angle and ${\bf n}$ is a unit vector indicating the direction of spin polarization; thus the FMC order parameter lives in manifold
\beq
M_{FMC}=U(1)\times S^2.
\label{FMorderparameterspace}
\eeq
Here $S^2$ (or 2-sphere) is the space of ${\bf n}$, $U(1)$ (also referred to as $S^1$ in the literature) is the space of $\theta$. The dimensionality of this order parameter space is thus $2+1=3$.
The ground state spin quantum number is $S_{tot}=N$, where $N$ is the atom number. So the ground state is fully magnetized.

(ii) Nematic condensate (NMC) phase with $0 < c_2 < c_2^*$, where $a_2 > a_0 > -\infty$. In this phase atoms condense into a nematic state:
$\zeta_{NMC}=e^{i\theta}(0, 1, 0)^T$,
up to an arbitrary global rotation and gauge transformation. The order parameter can be parameterized as a {\em product} ${\bf n}e^{i\theta}$, where $\theta$ is the superfluid phase angle and ${\bf n}$ is a unit vector indicating the direction of spin nematic. It is extremely important to identify $(\theta, {\bf n})$ and $(\theta+\pi, -{\bf n})$ as the same point in order parameter space\cite{zhou}, which is implied in our parameterization, ${\bf n}e^{i\theta}$. As a result the nematic spinor condensate order parameter lives in the coset manifold space\cite{zhou,moore}
\beq
M_{NMC}=[U(1)\times S^2]/Z_2,
\label{orderparameterspace}
\eeq
where the $Z_2$ reduction corresponds to the identification of $(\theta, {\bf n})$ with $(\theta+\pi, -{\bf n})$. The dimensionality of the order parameter is still 3.

It is clear that the boundary separating the FMC and NMC phases is at $c_2=0$, where $a_0=a_2$. This is a 1st order phase boundary across which magnetization changes by $N$.

(iii) As $c_2$ increases, the two-particle interaction turns attractive in the singlet channel (and $a_0$ turns negative), and beyond certain threshold bound states or singlet molecules can form (where $a_0$ turns back to be positive). In this case the ground state is a Bose condensate of singlet molecules, which has no spin structure\cite{cao}. This molecular, or pair condensate (PC) state breaks the U(1) symmetry associate with charge conservation, but respects the SU(2) spin rotation symmetry; its order parameter is
$e^{i\theta'}$, where $\theta'$ is the superfluid phase angle of the molecules. Thus the PC order parameter space is simply
\beq
M_{PC}=U(1),
\label{PCorderparameterspace}
\eeq
which is 1-dimensional. As a result there must be a {\em quantum} phase transition between the NMC state and the PC state at $T=0$, driven by increasing quantum fluctuation parameterized by $c_2$. We identify the QPT point as $c_2^*$, which turns out to be a quantum critical point (QCP), as we show below. In the {\em dilute} limit, $c_2^*$ corresponds to the Feshbach resonance in the singlet channel where $|a_0|\rightarrow \infty$.

To study the critical properties of this QCP, we write down the quantum (imaginary time) effective action that describes the NMC and PC phases on equal footing:
\beqarr
S&=&{\rho_{\bf n}\over 2}\int{d\tau d{\bf r}}\left[(\nabla{\bf n})^2 + {1\over v_{\bf n}^2} (\partial_\tau{\bf n})^2 \right]\nonumber\\
&+&{\rho_{\theta}\over 2}\int{d\tau d{\bf r}}\left[(\nabla{\theta})^2+ {1\over v_{\theta}^2} (\partial_\tau{\theta})^2\right]\nonumber\\
&+& {\rho_{\theta'}\over 2}\int{d\tau d{\bf r}}\left[(\nabla{\theta'})^2+ {1\over v_{\theta'}^2} (\partial_\tau{\theta'})^2\right]\nonumber\\
&+& J \int{d\tau d{\bf r}}\cos(2\theta-\theta')],
\label{effectiveaction}
\eeqarr
where the first two terms is the action for the NMC\cite{zhou}, the 3rd term is the action for the PC, and last term describes the conversion between two atoms and a singlet pair. Such ``two-channel" models have been used to study pairing induced QPT in spinless\cite{radzihovsky} and spin-1/2\cite{yang} bosons. The form of this effective action is essentially determined by the symmetry properties of these two phases, and their order parameter dynamics or collective mode spectra. We have neglected terms consistent with symmetry but involve more gradients, which are irrelevant in the long-distance and low-frequency limit.

It should be noted that the PC order parameter $\theta'$ is ordered in {\em both} the nematic and pair condensates states; it thus has a non-zero expectation value at the transition, with fluctuations that can be safely integrated out\cite{radzihovsky,yang}. Thus for the description of this transition, we can choose (without loss of generality) $\theta'=0$, and the last term of Eq. (\ref{effectiveaction}) reduces to
$J' \int{d\tau d{\bf r}}\cos(2\theta)$.
This reduces $\theta$ to an Ising or $Z_2$-like variable, as the expression above takes minima at $\theta=0$ and $\theta=\pi$\cite{zhouisingnote}. However because $(\theta, {\bf n})$ and $(\theta+\pi, -{\bf n})$ correspond to the same point in the NMC order parameter space, we have the freedom of making the ``gauge-fixing" choice of $\theta=0$; this in turn removes the redundancy in the space $U(1)\times S^2$, thus ${\bf n}$ now lives on the full $S^2$ space after this ``gauge-fixing" procedure.

After integrating out massive fluctuations of $\theta$, we arrive at the appropriate field theory for the QPT between NMC and PC phases:
\beq
S_{\bf n}={\rho'_{\bf n}\over 2}\int{d\tau d{\bf r}}\left[(\nabla{\bf n})^2 + {1\over {v'_{\bf n}}^2} (\partial_\tau{\bf n})^2 \right],
\eeq
where ${\bf n}$ is a unit vector of 3D spin space that lives in the {\em full} $S^2$ space. This is nothing but the O(3) non-linear $\sigma$ model (NL$\sigma$M). We thus conclude the universality class of the QPT at $c_2^*$ is that of $d+1$-dimensional O(3) or Heisenberg transition where $d=2$ or 3 is the spatial dimension of the system.

We note that this result is quite unusual, and the presence of superfluidity on {\em both} sides of the transition is of crucial importance. Normally one expects a 1st order transition between nematic and isotropic phases\cite{lubensky}, when no superfluidity is involved. In insulating quantum magnets a continuous transition is possible, but it is of a more exotic type\cite{senthil}.

In the theory (\ref{effectiveaction}) we have neglected the atom-molecule and molecule-molecule interactions. In the case of spinless bosonic atoms\cite{radzihovsky} such interactions are generally attractive, which render the PC phase unstable in that case. In the following we show this is {\em not} the case here. This is most easily demonstrated by going to the strong pairing limit $c_2\rightarrow\infty$, and use a lattice regularization in which the integrals in Eq. (\ref{hamiltonian}) are turned into summations over lattice sites. In this case the interaction terms in (\ref{hamiltonian}) take the form
\beq
V=\sum_{i} {[(c_0/2)\hat{n}_i(\hat{n}_i-1)+(c_2/2)(\hat{{\bf s}}_i^2-2\hat{n}_i)]},
\eeq
where $\hat{n}_i$ and $\hat{{\bf s}}_i$ are the total atom number and spin operators of site $i$ respectively. In the strong pairing limit the molecule is simply a singlet pair of two atoms on a given site, whose energy is $E_{pair}=c_0-2c_2$, while a single atom has energy $E_{atom}=0$. Now consider the interaction energy of a molecule and an atom occupying the same site (or equivalently, 3 atoms on one site with total spin $s=1$): $E_3=3c_0-2c_2 > E_{pair}+E_{atom}$. Similarly, consider two molecules on the same site (or 4 atoms forming a total singlet): $E_4=6c_0-4c_2 > 2E_{pair}$. We thus conclude that the atom-molecule and molecule-molecule interactions are both repulsive in the present case, and thus the PC phase is stable.

\begin{centering}
\begin{figure}
\epsfig {file=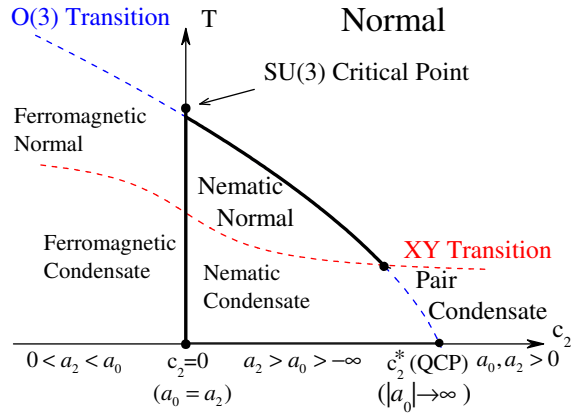,width=80mm}
\caption{(Color online)
Phase diagram of a spin-1 Bose gas in 3D. Thick solid lines are 1st order phase boundaries, while dashed lines are 2nd order phase boundaries, with universality classes indicated with the same color.
}
\label{fig2}
\end{figure}
\end{centering}

{\em Finite Temperature Phases and Phase Transitions in 2D} -- It is well-known that thermal fluctuations suppresses ferromagnetic or nematic spin order in 2D\cite{lubensky}. As a result among the three phases at $T=0$, only the PC phase survives at finite $T$ (with power-law long range order), which eventually yields to the normal phase via a KT transition. All these are reflected in the phase diagram, Fig. 1. It is interesting to note that within the PC phase and at very low $T$, as one goes from small to large $c_2$, there is a crossover similar to the BCS-BEC crossover in spin-1/2 fermionic superfluids, in the following sense. In this phase one can identify the pair size $\xi$ as correlation length of the spin nematic order parameter. At small $c_2$ we have $\xi\gg l$, where $l$ is the average interatomic distance, very much like what happens in a weak-coupling BCS fermionic superfluid. For large $c_2$ atoms form closely bound singlet molecules, thus $\xi$ is the size of the molecule, and we have $\xi \ll l$, similar to the BEC regime of the fermionic superfluid. The crossover happens near $c^*_2$.

{\em Finite Temperature Phases and Phase Transitions in 3D} -- The 3D phase diagram is considerably richer, as all three phases at $T=0$ are stable at low-$T$. Similar to the 2D case, the PC phase yields to the normal phase via a 3D XY (or O(2)) transition as $T$ increases. For $c_2 \lesssim c_2^*$ (in the NMC phase but near the QCP), we expect the nematic order parameter to be suppressed and the NMC phase first yields to the PC and then to normal phase as $T$ increases. Using analysis similar to
the QPT between NMC and PC phases, we conclude the {\em classical} NMC to PC transition is in the universality of 3D O(3) transition. The transition temperature $T_c$ near the QCP is determined by the critical properties of the QCP\cite{sachdevbook}:
\beq
T_c\propto (c^*_2-c_2)^{z\nu},
\label{tc}
\eeq
where in the present case the dynamical exponent $z=1$, and because $d+1=4$ is the upper critical dimension, the QCP correlation length exponent $\nu=1/2$ but there may be logarithmic correction to (\ref{tc}). For reasons to be elaborated below, we expect the O(3) and XY phase boundaries to cross at $c_2'$ with $0 < c_2' < c_2^*$, so for $0 < c_2 < c_2'$, we expect the NMC phases to first yield to a normal phase with nematic spin order via an XY transition, and then to the isotropic normal phase. The nematic to isotropic transition is expected to be 1st order\cite{lubensky}.

On the ferromagnetic side ($c_2 < 0$), condensates (including multiparticle condensates if present) must also have ferromagnetic spin order. Thus the order of transitions must be that the FMC phase first yields to a normal ferromagnetic phase via an XY transition, and then to the isotropic normal phase via an O(3) transition. This persists all the way to the SU(3) line of $c_2=0$. The separation of magnetic and superfluid transition temperatures can be easily established by a perturbative treatment of $c_0$ and $c_2$.

Due to the enhanced symmetry, along the $c_2=0$ line the order parameter lives in a much larger space, which is a 5-dimensional manifold:
\beq
M_{c_2=0}=U(1)\times \{SU(3)/[U(1)\times SU(2)]\},
\eeq
where the first factor $U(1)$ is the same as in previous cases, while the 2nd factor represents the 3-component spinor order parameter $\phi$ with the constraint $\phi^\dagger\phi=1$ and excluding its overall phase; the SU(2) reduction there is the subgroup of SU(3) that leaves $\phi$ invariant.

For the same reason as in the $c_2 < 0$ case, as $T$ increases, we expect the U(1) symmetry to be restored first, and then the SU(3) symmetry. The effective action for the SU(3) critical point is the CP2 model:
\beq
S[\phi]={\rho_\phi\over 2}\int{d{\bf r}}[(\nabla+i{\bf A})\phi]^\dagger\cdot[(\nabla+i{\bf A})\phi].
\eeq
The high symmetry also guarantees the $c_2=0$ line is a 1st order boundary separating ferromagnetic and nematic phases, with or without condensates. Furthermore, the XY transition line and isotropic to ferromagnetc or nematic transition line must be continuous at $c_2=0$. As a result for small positive $c_2$ there must be a nematic normal phase separating the NMC and isotropic normal phases. All these are summarized in the 3D phase diagram, Fig. 2.

{\em Phases in 1D} -- Due to strong thermal fluctuations, the only stable phase at finite $T$ is the isotropic normal phase. At $T=0$ the NMC phase is destroyed by quantum fluctuations, thus only the FMC and PC phases remain, separated by the 1st order QPT at $c_2=0$. The ferromagnetic order parameter is a conserved quantity and thus not affected by the enhanced quantum fluctation, but the superfluid order parameters only have power-law long-range order in both phases.

The author thanks Jason Ho, Joel Moore, Subir Sachdev, Fei Zhou and in particular Ashvin Vishwanath for helpful conversations, and W. Ding for graphics assistance. This work was supported by National Science Foundation grant No. DMR-0704133, and completed while the author was visiting Kavli Institute for Theoretical Physics (KITP). The work at KITP was supported in part by National Science Foundation grant No. PHY-0551164.

\end{document}